\documentclass{article}

\usepackage{PRIMEarxiv}

\usepackage{algorithm}
\usepackage{algpseudocode}
\usepackage{xcolor, colortbl}
\usepackage{listings}
\usepackage{balance}
\usepackage{subcaption}
\usepackage{graphicx}
\usepackage{comment}
\usepackage{amsmath}

\usepackage{hyperref}       
\usepackage{booktabs}       
\usepackage{nicefrac}       
\usepackage{microtype}      
\usepackage{lipsum}
\usepackage{fancyhdr}       

\newcommand{\maq}{\texttt}

\newcommand{\ita}{\textit}
\newcommand{\ngr}{\textbf}

\newcommand{\iit}{\begin{itemize}}
\newcommand{\fit}{\end{itemize}}
\newcommand{\ienle}{\begin{enumerate}[(a)]}
\newcommand{\ien}{\begin{enumerate}}
\newcommand{\fen}{\end{enumerate}}

\definecolor{codeOrange}{rgb}{1.0,0.5,0.0}
\definecolor{codeGreen}{rgb}{0,0.5,0}
\definecolor{codeGray}{rgb}{0.5,0.5,0.5}
\definecolor{codeBlue}{rgb}{0.1,0.3,1.0}
\definecolor{codeRed}{rgb}{1.0,0.3,0.2}
\title{Deep learning techniques for blind image super-resolution: A high-scale multi-domain perspective evaluation
}

\author{
 Valdivino Alexandre de Santiago J\'unior \\
 Coordena\c{c}\~ao de Pesquisa Aplicada e Desenvolvimento Tecnol\'ogico (COPDT) \\
 Instituto Nacional de Pesquisas Espaciais (INPE)  \\ 
 S\~ao Jos\'e dos Campos, S\~ao Paulo, Brazil \\
 \texttt{valdivino.santiago@inpe.br} \\ }

\begin{document}
\maketitle


\begin{abstract}
Despite several solutions and experiments have been conducted recently addressing image super-resolution (SR), boosted by deep learning (DL) techniques, they do not usually design evaluations with high scaling factors, capping it at 2x or 4x. Moreover, the datasets are generally benchmarks which do not truly encompass significant diversity of domains to proper evaluate the techniques. It is also interesting to remark that blind SR is attractive for real-world scenarios since it is based on the idea that the degradation process is unknown, and hence techniques in this context rely basically on low-resolution (LR) images. In this article, we present a high-scale (8x) controlled experiment which evaluates five recent DL techniques tailored for blind image SR: Adaptive Pseudo Augmentation (APA), Blind Image SR with Spatially Variant Degradations (BlindSR), Deep Alternating Network (DAN), FastGAN, and Mixture of Experts Super-Resolution (MoESR). We consider 14 small datasets from five different broader domains which are: aerial, fauna, flora, medical, and satellite. Another distinctive characteristic of our evaluation is that some of the DL approaches were designed for single-image SR but others not. Two no-reference metrics were selected, being the classical natural image quality evaluator (NIQE) and the recent transformer-based multi-dimension attention network for no-reference image quality assessment (MANIQA) score, to assess the techniques. Overall, MoESR can be regarded as the best solution although the perceptual quality of the created HR images of all the techniques still needs to improve. We believe that independent and unbiased evaluations, like this one, are important to indicate the most suitable approaches to be selected by professionals who need to work with image SR in their practical settings. Supporting code: \href{https://github.com/vsantjr/DL_BlindSR}{https://github.com/vsantjr/DL\_BlindSR}. Datasets: \href{https://www.kaggle.com/datasets/valdivinosantiago/dl-blindsr-datasets}{https://www.kaggle.com/datasets/valdivinosantiago/dl-blindsr-datasets}.
\end{abstract}



\section{Introduction} \label{intro}

Medical imaging \cite{medicalsr0, medicalsr1, medicalsr2, medicalsr3}, internet video delivery \cite{videodeliverysr0, videodeliverysr1, videodeliverysr2}, surveillance and security via person identification \cite{survface0,survface1,survface2}, and remote sensing \cite{rssrsurvey, sreegan, srddrn, srtesagan, srhan, srrban} are just some examples of real-world applications where image super-resolution (SR) has been used. In SR, we aim at recovering high-resolution\footnote{In this article, image resolution means precisely the dimensionality of the image. For instance, an image has a resolution of $W \times H$ pixels. This is not to be confused with other definitions of resolution existing in certain communities such as remote sensing (spatial resolution, radiometric resolution, temporal resolution, ...).} (HR) images from low-resolution (LR) ones. This is a non-trivial and generally ill-posed problem since multiple HR images exist corresponding to a unique LR image \cite{surveysr}.

Several classical methods have been designed for image SR such as bicubic interpolation and Lanczos resampling \cite{lanczos/79}, edge-based methods \cite{jianedge/08}, statistical methods \cite{xiongstat/10}, among others. But, naturally, with all the developments related to deep learning (DL) and deep neural networks (DNNs) \cite{dlmeta/21}, a significant number of strategies have been proposed addressing SR via DL/DNNs as reported in recent secondary studies \cite{surveysr, rssrsurvey}. Among the DL techniques, convolutional neural networks (CNNs) \cite{cnndongsr/14,cnnkimsr/16,cnnlimsr/17,srddrn}, generative adversarial networks (GAN) \cite{srgan/17,esrgan/18, sreegan, srtesagan}, and attention-based networks \cite{attzhangsr/18, attchengsr/18, srhan} have been employed to solve image SR problems.

The majority of studies published up to now focus on supervised SR where models are trained with both LR images and the corresponding HR ones \cite{surveysr, rssrsurvey}. But, in reality, it is not easy to have images from the same scene but with distinct resolutions so that having the pairs LR-HR for training is not as direct as it is supposed to be. Hence, in unsupervised SR \cite{bulat/18}, only unpaired LR-HR images are available for training, so that the models are more able to learn real-world LR-HR mappings.

But an even more attractive strategy for real-world scenarios is blind image SR \cite{blindgu/19,blindyu/22} which is based on the idea that the degradation process/kernels is/are unknown, and hence techniques in this context rely basically on LR images, not requiring the high-resolution ones. There is an increasing interest in blind image SR.

Despite the huge number of proposed techniques and experiments boosted by DL techniques \cite{surveysr, rssrsurvey},  they do not usually design evaluations with high scaling factors, capping it at 2x or 4x. One of the exceptions is the experiment presented in \cite{rssrsurvey} where authors considered 2x, 4x, and 8x scaling factors. However, the authors considered a multi-sensor remote sensing dataset (MSRSD) consisting of mostly publicly available very HR satellite images. Even if the images are from different satellites and regions, eventually the diversity of the images and feature spaces are not enough to proper evaluate SR techniques. 

Several other datasets have been used for training image SR approaches such as BSDS300 \cite{bsds300}, BSDS500 \cite{bsds500}, DIV2K \cite{div2k}, PIRM \cite{pirm}, Set5 \cite{set5}, Set14 \cite{set14}, and Urban100 \cite{urban100}. These datasets comprise different categories but still lack of images of other domains. For instance, all the datasets above do not seem to present images obtained via satellite sensors or medical images. Interesting to stress that images taken by sensors embedded in satellites or airplanes have considerable differences compared to natural images due to different shooting content and shooting methods \cite{p26_cnn_metale}. Moreover, as previously remarked, evaluations are usually limited to the 4x scaling factor. In \cite{haris/18}, results are presented for 8x scaling factor but considering only the Set5 and Set14 datasets. We believe that performing experiments addressing not only high scaling factors (e.g. 8x) but also datasets of different domains is very important to better identify the most adequate image SR approaches.

Image quality assessment (IQA) metrics (methods) are roughly divided into two categories: full-reference (FR-IQA) and no-reference (NR-IQA) \cite{ntire/22}. In FR-IQA, we evaluate the similarity between a distorted image and a given reference image, and classical metrics which have been extensively used are peak signal-to-noise ratio (PSNR) \cite{rssrsurvey} and structural similarity index measure (SSIM) \cite{structsim/04}. On the other hand, NR-IQA metrics are proposed to assess image quality without a reference
image, being more suitable for perceptual quality. Natural image quality evaluator (NIQE) \cite{niqe} and perception index (PI) \cite{pirm} are two traditional metrics. However, recent metrics can be devised based on DNNs as presented in the \ngr{New Trends in Image Restoration and Enhancement} (NTIRE) workshop held at CVPR 2022 \cite{ntire/22}. And, for the first time, NTIRE 2022 Challenge on Perceptual Image Quality Assessment had a track addressing NR-IQA. We believe that for evaluating approaches in a completely ``blind" setting, NR-IQA metrics are more interesting since they do not demand a reference image, only the LR ones.

In this article, we present a high-scale (8x) controlled experiment which evaluates five recent DL techniques tailored for blind image SR: Adaptive Pseudo Augmentation (APA) \cite{apa/21}, Blind Image SR with Spatially Variant Degradations (BlindSR) \cite{blindsrsel/19}, Deep Alternating Network (DAN) \cite{dan/20}, FastGAN \cite{fastgan/21}, and Mixture of Experts Super-Resolution (MoESR) \cite{moesr/22}. Relying basically on public sources, we adapt and create 14 small (100 samples) LR images datasets from five different broader domains: aerial, fauna, flora, medical, and satellite\footnote{The aerial broader domain means data/images obtained at altitudes lower than 100 km (62 miles; K{\'a}rm{\'a}n Line) above the mean sea level. For example, images obtained by sensors attached/embedded in airplanes or unmanned aerial vehicles (drones). The satellite broader domain relates to the space and it implies data/images obtained from sensors, usually onboard satellites, which are at a minimum altitude of 100 km above the mean sea level.}. Another distinctive characteristic of our evaluation is that some of the DL approaches were designed for single-image SR but others not. 

Two NR-IQA metrics were selected, being the classical NIQE and the recent vision transformer(ViT)-based multi-dimension attention network for no-reference image quality assessment (MANIQA) score \cite{maniqa/22}, to assess the techniques. The MANIQA model was the winner of the NTIRE 2022's NR-IQA track obtaining a performance considerably higher than classical strategies such as PI and NIQE.

The contributions of this study are:

\ien
\item We design and execute a controlled experiment considering five recent blind image SR approaches and focusing on a specific large scaling factor (8x). Note that we decided to present a more detailed analysis of the results, trying to explain in more depth the behaviours of the approaches. We also perform a correlation analysis taking into account the HR images produced by the two best DL techniques. We believe that independent and unbiased evaluations are important to indicate the most suitable approaches to be selected by professionals who need to work with image SR in their practical settings; 
\item We adapt and create 14 small (100 samples) LR images datasets from five different broader domains: aerial, fauna, flora, medical, and satellite. We believe that making available these small datasets to the community, obtained from quite distinct domains, is interesting to provide other possibilities to evaluate the blind image SR techniques considering not only single-image but non-single-image techniques (we had indeed done this);
\item We consider a recent DNN-based NR-IQA metric (MANIQA score) in addition to a classical one (NIQE), and present some remarks by using such metrics.
\fen

This article is structured as follows. Section \ref{background} briefly presents the theoretical background and related work. In Section \ref{expdesign}, we show in detail the design of our experiment. Results are in Section \ref{results} while Section \ref{disc} discusses some important points. In Section \ref{conc}, conclusions and feature directions are pointed out.

\section{Background} \label{background}

This section presents an overview of the theoretical background related to image SR. More details can be seen elsewhere \cite{surveysr, rssrsurvey}. At the end, we also discuss about related work. The goal of image SR is to recover the corresponding HR images from the LR images. The LR image, $I_{LR}$, is usually modelled as the output of the following degradation:

\begin{equation}
I_{LR} = \Gamma(I_{HR}; \delta) 
\end{equation}

\noindent where $\Gamma$ means a degradation mapping function, $I_{HR}$ is the corresponding HR image, and $\delta$ represents the parameters of the degradation process. When the degradation process ($\Gamma$ and $\delta$) is unknown, which is very common to happen, and only LR images are available for the techniques, we have the so called blind image SR. In this case, the idea is to recover an HR approximation, $\widehat{I_{HR}}$, of the ground truth HR image, $I_{HR}$, from the LR image as presented below:

\begin{equation}
\widehat{I_{HR}} = \mathcal{M}(I_{LR}; \theta) 
\end{equation}

\noindent where $\mathcal{M}$ is the SR model and $\theta$ are the parameters of $\mathcal{M}$.

Despite of the fact that the degradation process is usually unknown, several studies model the degradation as follows:

\begin{equation}
\Gamma(I_{HR}; \delta) = (I_{HR}) \downarrow s, \{s\} \subset \delta 
\end{equation}

\noindent where $\downarrow$ is a downsampling operation and $s$ is a scaling factor. But, some studies \cite{zhanglearn/18} do such a degradation modelling as presented below:

\begin{equation}
\Gamma(I_{HR}; \delta) = (I_{HR} \otimes \kappa) \downarrow s + \varphi_\xi, \{\kappa, s, \xi\} \subset \delta 
\end{equation}

\noindent where $I_{HR} \otimes \kappa$ means the convolution between a blur kernel, $\kappa$, and the HR image, $I_{HR}$, and $s$ is a scaling factor. Moreover, $\varphi_\xi$ is some additive white Gaussian noise with standard deviation $\xi$.  

\subsection{Metrics}

As previously mentioned, there are several metrics for IQA, being full-reference (FR-IQA) or without relying on a reference (NR-IQA). We present here a brief discussion about the two NR-IQA metrics we have considered in this research: NIQE \cite{niqe} and MANIQA score \cite{maniqa/22}.

The main motivation to develop NIQE is that, at the time it was proposed, NR-IQA models required knowledge about anticipated distortions in the form of training samples and corresponding human opinion scores. Thus, NIQE is a completely blind approach which only uses measurable deviations from statistical regularities observed in natural images, without training on human-rated distorted images and, hence, without any exposure to distorted images. The idea is to construct a ``quality aware" collection of statistical features based on a successful space domain natural scene statistic (NSS) model. 

NIQE is derived by computing 36 identical NSS features from patches of the same size, $P \times P$, from the image to be analysed, fitting them with a multivariate Gaussian model (MVG) \cite{mgauss/12}, then comparing its MVG fit to the natural MVG model. They considered a patch of dimension (resolution) $96 \times 96$, although other dimensions were also evaluated. The quality of the distorted image is expressed as the distance between the quality aware NSS feature model and the MVG fit to the features extracted from the distorted image, as shown below:

\begin{equation}
\mathcal{D}(\nu_1,\nu_2,\Sigma_1,\Sigma_2)=\sqrt{\Bigg ( (\nu_1 - \nu_2)^T \Bigg (\frac{\Sigma_1 + \Sigma_2}{2}\Bigg )^{-1} (\nu_1 - \nu_2) \Bigg )}
\end{equation}

\noindent where $\nu_1$ and $\nu_2$ are the mean vectors of the natural MVG model and the distorted image’s MVG model, respectively. Furthermore, $\Sigma_1$ and $\Sigma_2$ are the covariance matrices of the natural MVG model and the distorted image’s MVG model, respectively. Thus, as \ngr{lower} the NIQE, the better the perceptual quality of the generated HR image.

\begin{figure}[!htb]
\centering
\includegraphics[width=1.0\textwidth]{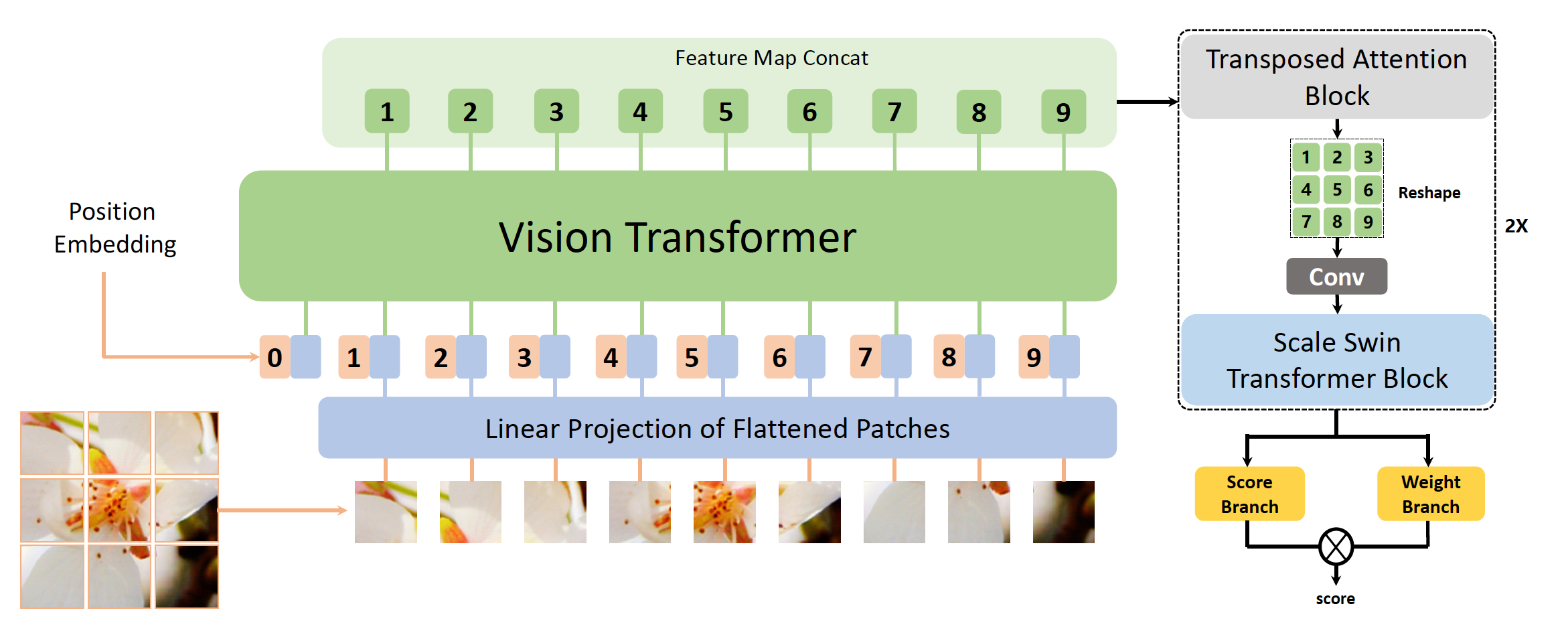}
\caption{The MANIQA model. The score is the metric related to the image. Source: adapted from \cite{maniqa/22}}
\label{maniqafig}
\end{figure}

While NIQE is a very traditional metric, the MANIQA score is a very recent one. Top approach of the NTIRE 2022's NR-IQA track \cite{ntire/22}, the motivation for the development of the MANIQA model and its corresponding score is that NR-IQA metrics/methods are limited when assessing images created by GAN-based image restoration algorithms \cite{esrgan/18,ganir/20}. As shown in Fig. \ref{maniqafig}, the MANIQA model consists of four components: feature extractor using ViT \cite{vit/21}, transposed attention block, scale swin transformer block, and a dual-branch structure for patch-weighted quality prediction.

In more detail, a distorted image is cropped into patches of dimension 8 x 8. Then, the patches are inputted into the ViT for extracting the features. Transposed attention block and scale swin transformer block are used to strengthen the global and local interaction. A dual-branch structure is proposed for predicting the weight and score of each patch. Notice that as \ngr{higher} the MANIQA score, the better the perceptual quality of the generated HR image. 

\subsection{Related work}

This study is an experiment within blind image SR via DL/DNNs. Basically every study within blind image SR presents experimental evaluations and, in lesser or greater extent, they are related to ours. As mentioned in Section \ref{intro}, even if there are so many techniques and experiments proposed boosted by DL techniques, as corroborated by secondary studies \cite{surveysr, rssrsurvey}, in general the studies do not consider high scaling factors such as 8x. When there are exceptions  \cite{rssrsurvey,haris/18}, there is no significant diversity of images and feature spaces to proper evaluate the image SR approaches, taking into account quite distinct broader domains such as medicine images, images obtained by satellites via sensors with different characteristics, and images more ``usual'' like those of animal's faces. 

It is worth noting that dealing with this range of different domains is important to better classify the techniques in terms of their performances. In other words, the wider/general the assessments are, the better. Furthermore, independent evaluations, like the one presented in this article, are valuable to guide professionals in making the correct decisions in choosing the most appropriate blind image SR solution. Our study addresses all these previous points and these are the main differences between our effort and other already published articles.

\section{Experiment design} \label{expdesign}

In this section, we describe the main design options of the controlled experiment to assess the performance of the five DL techniques for blind image SR: APA, BlindSR, DAN, FastGAN, and MoESR. 

\subsection{Research questions and variables}

The research questions (RQs) related to this experiment are: 
\ien
\item \textbf{RQ\_1} - Which out of the five algorithms for blind image SR is the best regarding the metrics NIQE and MANIQA score? And which can be considered the best overall? 
\item \textbf{RQ\_2} - Does the two top approaches present similar behaviours when deriving HR images?

\fen

The motivation for \textbf{RQ\_1} is self-explained. Regarding \textbf{RQ\_2}, the idea here is to perceive whether the two best approaches present similar behaviours. In other words, are the images detected as having the best, as well as the worst, quality perceptions, based on the MANIQA scores, somewhat ``common" to both best DL techniques? Our goal here is to see whether the best algorithms ``agree" in terms of the HR images they create based only on the LR ones. The independent variables are the DL models. The dependent variables are the values of the metrics, i.e. NIQE and MANIQA score. 

\subsection{Datasets}

The 14 datasets comprise five different broader domains: aerial, fauna, flora, medical, and satellite. Tables \ref{datasets1} and \ref{datasets2} present details about the datasets we created based on publicly released images. In Table \ref{datasets2}, note that the original resolution of the images in the datasets are diverse. Some source datasets, like \texttt{condoaerial} and \texttt{massachbuildings}, are formed by HR images where all samples have the same resolution. The \texttt{plantpat} dataset has HR images too but there are several different resolutions. Others like \texttt{catsfaces} and \texttt{isaid} have HR, LR and even medium resolution images. Three Satellite datasets, \texttt{amazonia1}, \texttt{cbers4a}, and \texttt{deepglobe}, are composed only by LR images and all have the same resolution.

\begin{table}[!htb]
\centering
\caption{Datasets: description and source}
\begin{tabular}{c|c|p{3.5cm}|p{3.2cm}}
\hline
\footnotesize \ngr{Domain} & \footnotesize \ngr{Dataset} & \footnotesize \ngr{Description} & \footnotesize \ngr{Source} \\ 
\hline
 
\footnotesize Aerial & \footnotesize \maq{condoaerial} &  \footnotesize Aerial Semantic Segmentation Drone Dataset & \footnotesize \href{https://bit.ly/3YWXNYy}{https://bit.ly/3YWXNYy} \\ 
 & \footnotesize   \maq{massachbuildings} & \footnotesize Massachusetts Buildings Dataset & \footnotesize \href{https://bit.ly/41nS4fI}{https://bit.ly/41nS4fI} \\ 
 & \footnotesize   \maq{ships} & \footnotesize Ship Detection from Aerial Images Dataset & \footnotesize \href{https://bit.ly/3ZaPEiK}{https://bit.ly/3ZaPEiK} \\ 
 & \footnotesize   \maq{ufsm-flame} & \footnotesize Drone Images from UFSM and Flame Datasets & \footnotesize \href{https://bit.ly/3KxJzbZ}{https://bit.ly/3KxJzbZ} \href{https://bit.ly/3YUwA8B}{https://bit.ly/3YUwA8B} \href{https://bit.ly/3ZhiSMZ}{https://bit.ly/3ZhiSMZ} \\   \hline
 \hline
 
 \footnotesize Fauna & \footnotesize \maq{catsfaces} & \footnotesize Cats Faces Dataset & \footnotesize \href{https://bit.ly/41gDrLv}{https://bit.ly/41gDrLv} \\ 
 & \footnotesize   \maq{dogsfaces} & \footnotesize Dogs Faces Dataset & \footnotesize \href{https://bit.ly/41gDrLv}{https://bit.ly/41gDrLv} \\   \hline
 \hline
 
 \footnotesize Flora & \footnotesize \maq{flowers} & \footnotesize  102 Category Flower Dataset & \footnotesize \href{https://bit.ly/3EAxH5w}{https://bit.ly/3EAxH5w} \\ 
 & \footnotesize  \maq{plantpat} & \footnotesize Plant Pathology 2021 - FGVC8 - Dataset & \footnotesize \href{https://bit.ly/3EBCkMv}{https://bit.ly/3EBCkMv} \\   \hline
 \hline
 
 \footnotesize Medical & \footnotesize \maq{melanomaisic} & \footnotesize SIIM-ISIC Melanoma Classification Dataset & \footnotesize \href{https://bit.ly/3XUqELO}{https://bit.ly/3XUqELO} \href{https://bit.ly/3m0Az5b}{https://bit.ly/3m0Az5b} \\ 
 & \footnotesize  \maq{structretina} & \footnotesize Structured Analysis of the Retina Dataset & \footnotesize \href{https://bit.ly/3ZjfzVJ}{https://bit.ly/3ZjfzVJ} \\  \hline
 \hline
 
 \footnotesize Satellite & \footnotesize \maq{amazonia1} & \footnotesize Cloudless Scene from Amazonia 1 Satellite Dataset & \footnotesize \href{https://bit.ly/3m3qKTP}{https://bit.ly/3m3qKTP} \\ 
 & \footnotesize \maq{cbers4a} & \footnotesize Scene with Clouds from CBERS-4A Satellite Dataset & \footnotesize \href{https://bit.ly/3m3qKTP}{https://bit.ly/3m3qKTP} \\ 
 & \footnotesize \maq{deepglobe} & \footnotesize Forest Aerial Images for Segmentation Dataset & \footnotesize \href{https://bit.ly/3Eu4Nnq}{https://bit.ly/3Eu4Nnq} \\ 
 & \footnotesize \maq{isaid} & \footnotesize Instance Segmentation in Aerial Images Dataset & \footnotesize \href{https://bit.ly/3IpavYU}{https://bit.ly/3IpavYU} \\ 
 \hline

\end{tabular}
\label{datasets1}
\end{table}

\begin{figure}[!htb]       
    \includegraphics[width=0.16\linewidth]{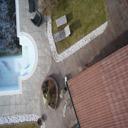}
    \includegraphics[width=0.16\linewidth]{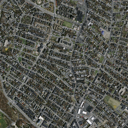}
    \includegraphics[width=0.16\linewidth]{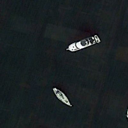}
    \includegraphics[width=0.16\linewidth]{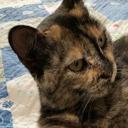}
    \includegraphics[width=0.16\linewidth]{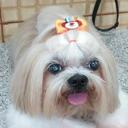}
    \includegraphics[width=0.16\linewidth]{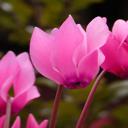}
    
     \includegraphics[width=0.16\linewidth]{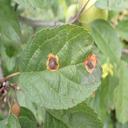}
    \includegraphics[width=0.16\linewidth]{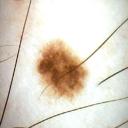}
    \includegraphics[width=0.16\linewidth]{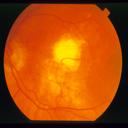}
    \includegraphics[width=0.16\linewidth]{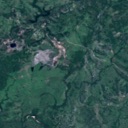}
    \includegraphics[width=0.16\linewidth]{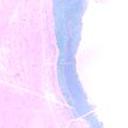}
    \includegraphics[width=0.16\linewidth]{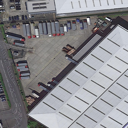}

     \caption{LR images from our datasets. Top row, left to right: \maq{condoaerial}, \maq{massachbuildings}, \maq{ships}, \maq{catsfaces}, \maq{dogsfaces}, \maq{flowers}; Bottom row, left to right: \maq{plantpat}, \maq{melanomaisic}, \maq{structretina}, \maq{amazonia1}, \maq{cbers4a}, \maq{isaid} }
    \label{samples}
\end{figure}

Given the different domains and image resolutions (in some cases, as shown in Table \ref{datasets2}, we just had at our disposal LR images), we therefore decided to default the input LR images to a resolution of 128 x 128 pixels. Thus, we downsampled the images which have higher resolution than that based on the bicubic interpolation method as many others have been doing \cite{surveysr, rssrsurvey}. Fig. \ref{samples} presents some LR images from the datasets we created.

Note that some criticise doing such a resizing because real-world cameras actually accomplishes a series of operations (demosaicing, denoising, ...) to finally produce 8-bit RGB images \cite{surveysr}. Thus, RGB images have lost lots of original signals becoming significant different from the original images taken by the camera. Therefore, it would not be interesting to directly use the manually downsampled RGB images by using, for instance, a bicubic interpolation method for image SR.

However, we must emphasise that our main goal is to perform an experiment using DL techniques for blind image SR, based on different domains, in order to provide some sort of recommendation to professionals who need to choose a technique that best fit their needs. Although the issues cited earlier exist, they impact equally all the DL techniques and hence they do not compromise our analysis. 

In addition to standardising the resolution of the LR input images, we also considered the same small number of images in the datasets: 100 samples. More details about this point in the next section.

\begin{table}[!htb]
\centering
\caption{Datasets: original resolution of the images}
\begin{tabular}{c|c|p{6.7cm}}
\hline
\footnotesize \ngr{Domain} & \footnotesize \ngr{Dataset} & \footnotesize \ngr{Original Resolution ($W  \times H$)} \\ 
\hline
 
\footnotesize Aerial & \footnotesize \maq{condoaerial} &  \footnotesize 6000 x 4000 \\ 
 & \footnotesize   \maq{massachbuildings} & \footnotesize 1500 x 1500 \\ 
 & \footnotesize   \maq{ships} & \footnotesize 568 x 526, 556 x 528, 550 x 570, …, 294 x 244, 266 x 254, 238 x 232 \\ 
 & \footnotesize   \maq{ufsm-flame} & \footnotesize 4000 x 3000, 254 x 254 \\   \hline
 \hline
 
 \footnotesize Fauna & \footnotesize \maq{catsfaces} & \footnotesize 2954 x 3027, 2893 x 3016, 2418 x 2161, …, 1024 x 837, 647 x 690, 256 x 256 \\ 
 & \footnotesize   \maq{dogsfaces} & \footnotesize 678 x 796, 256 x 256 \\   \hline
 \hline
 
 \footnotesize Flora & \footnotesize \maq{flowers} & \footnotesize  828 x 500, 819 x 500, 764 x 500, …, 500 x 525, 500 x 507, 500 x 500 \\ 
 & \footnotesize  \maq{plantpat} & \footnotesize  5184 x 3456, 4608 x 3456, 4032 x 3024, 4000 x 3000, 4000 x 2672, 3024 x 4032, 2592 x 1728 \\   \hline
 \hline
 
 \footnotesize Medical & \footnotesize \maq{melanomaisic} & \footnotesize 6000 x 4000, 5184 x 3456, 4288 x 2848, …, 1920 x 1080, 768 x 576, 640 x 480 \\ 
 & \footnotesize  \maq{structretina} & \footnotesize  700 x 605 \\  \hline
 \hline
 
 \footnotesize Satellite & \footnotesize \maq{amazonia1} & \footnotesize 128 x 128 \\ 
 & \footnotesize \maq{cbers4a} & \footnotesize 128 x 128 \\ 
 & \footnotesize \maq{deepglobe} & \footnotesize  256 x 256 \\ 
 & \footnotesize \maq{isaid} & \footnotesize 6661 x 6308, 6471 x 4479, 5963 x 5553, …, 436 x 793, 395 x 590, 211 x 521 \\ 
 \hline

\end{tabular}
\label{datasets2}
\end{table}

\subsection{Amount of required images}

Taking into account the ``amount" of required images, recent DL techniques for image SR are created assuming a single image, known as single-image SR, but they can also be devised assuming that there are a few (limited) number of samples, which we can call few-shot image SR. There is currently a great interest in few-shot learning (FSL) where we face the situation to solve a task based on few samples with supervised information \cite{fewshotsurvey/20}. Note that here we consider that few-shot SR assumes few but more than one image. There are also other possibilities, like zero-shot SR presented by \cite{zeroshot}, where authors coped with unsupervised SR by training image-specific SR networks at test time rather than training a model on huge datasets.  

Traditionally when researchers propose a new single-image SR technique, they naturally compare their approach to other single-image SR strategies. However, it would be nice to compare single-image SR techniques to non-single-image ones to perceive whether it is advantageous to consider approaches which rely on a unique image or, eventually, it is better to adopt solutions which require more than one image but not so many of them.

Thus, three of the selected techniques were designed addressing single-image SR, i.e. BlindSR, DAN, and MoESR. The other two, APA and FastGAN, are GAN-based approaches requiring a few samples indeed and were not conceived for single-image SR. Since the two latter did require more than one sample but not too many, we limited the size of the datasets to 100 LR images as mentioned in the previous section. In Section \ref{dltech}, we provide an overview of such DL strategies.

All runnings were performed using a Bull Sequana X1120 computing node of the SDumont supercomputer\footnote{\href{https://sdumont.lncc.br/machine.php?pg=machine}{https://sdumont.lncc.br/machine.php?pg=machine}}, where such a node has 4 x NVIDIA Volta V100 graphics processing units (GPUs). Each run was limited to four days being considered the latest model when the execution exceeded this time.

\subsection{DL techniques} \label{dltech}

In this section, we briefly describe the selected DL techniques starting with the single-image approaches. In \cite{blindsrsel/19}, authors proposed a framework that can achieve blind image SR in a fully automated manner, while also meeting the practical scaling needs of video production. We name it here as BlindSR and such a framework is composed of three main components. The first one is a degradation-aware SR network used to generate an HR image based on a LR input image and the corresponding blur kernel. Secondly, a kernel discriminator is trained to analyse the output HR image and predict errors caused by incorrect blur kernel input. Finally, an optimisation procedure aims to recover both the degradation kernel and the HR image by minimising the predicted error using their kernel discriminator. 

DAN solves the blind image SR problem via an alternating optimization algorithm which restores an HR image and estimates the corresponding blur kernel alternately \cite{dan/20}. Thus, these two subproblems are handled both via convolutional neural modules namely Restorer and Estimator, respectively. More specifically, the Restorer restores an HR image based on the predicted Estimator's kernel, and the Estimator estimates a blur kernel with the help of the restored HR image.

Some blind SR techniques train a unique degradation-aware network (for multiple kernels) on external datasets like BlindSR described above \cite{blindsrsel/19}. But there are performance problems with these approaches. Other direction are self-supervised techniques \cite{selfblind} but they are usually costly and there is limited information to learn from a single image. Aiming at benefiting of the best characteristics of both types of solution, MoESR was proposed considering different experts for different degradation kernels. For every input image, the technique predicts the degradation kernel and super-resolve the LR image using the most adequate kernel-specific expert. To predict the degradation kernel, they use two components in combination. Image Sharpness Evaluator (ISE) assesses the sharpness of the images generated by the experts. These evaluations are used by the Kernel Estimation Network (KEN) to estimate the kernel and select the best pretrained expert network.

Regarding the non-single-image techniques, the greatest motivation to create APA \cite{apa/21} is related to the main challenge in training GANs with limited (few) data: the risk of discriminator overfitting which can lead to unstable training dynamics \cite{limitedgan1,limitedgan2}. To deal with this issue, the technique called Adaptive Pseudo Augmentation, or APA, regularises the discriminator without introducing any external augmentations or regularisation terms. Unlike previous approaches that relied on standard data augmentations \cite{limitedgan1,limitedgan2}, APA leverages the generator within the GAN itself to provide the augmentation, a more natural way of regularisation of the overfitting of the discriminator. In accordance with the authors, the APA approach is simple and more adaptable to different settings and training conditions than model regularisation, without requiring manual tuning.

Other FSL solution and based on GANs is FastGAN \cite{fastgan/21}. Its authors were driven by the realisation that training GANs on high-fidelity (HR) images often necessitates a vast number of training images and large-scale GPU clusters. To address the challenge of few-shot image synthesis via GAN with minimal computing costs, FastGAN was proposed as a lightweight GAN architecture which produces high-quality results at a resolution of 1024 x 1024 pixels. Their technique involves incorporating two techniques: a skip-layer channel-wise excitation module and a self-supervised discriminator trained as a feature-encoder.

\subsection{Metrics}
As we have already mentioned, NIQE and the MANIQA score were the selected metrics. NIQE was calculated considering the HR images (1024 x 1024 pixels) generated by the techniques. However, according to its authors, the MANIQA model \cite{maniqa/22} is best suited for 222 x 224 pixel images. Thus, we downsampled the created HR image (1024 x 1024 pixels) to a resolution of 224 x 224 pixels in order to calculate the MANIQA score. 

Note that such a degradation approach accomplished to calculate the MANIQA score was the same for all generated HR images and, therefore, there is no favouring of a certain technique in relation to the obtained value. Furthermore, we got the MANIQA score considering a subset of the datasets and HR images (1024 x 1024) and we did not notice a great difference in the scores compared to when we used the reduced images (224 x 224). Thus, we followed the authors' suggestions and considered the images in the lowest resolution (224 x 224) as input for calculating the MANIQA score.

As shown in Table \ref{datasets2} and as we have already pointed out, the original images of some datasets are HR ones. But there are datasets which have only original LR images and not the corresponding HR samples. In addition to that, NR-IQA metrics are more in line with the reality when we deal with a blind setting. These are the reasons to select only NR-IQA metrics for our experiment.

\section{Results} \label{results}

In this section, we present the results of our experiment splitting the sections based on the research questions we have defined earlier.

\subsection{RQ\_1}

\subsubsection{NIQE}

\begin{figure}[p] 
  \begin{subfigure}[b]{0.5\linewidth}
    \centering
    \includegraphics[width=1.0\linewidth]{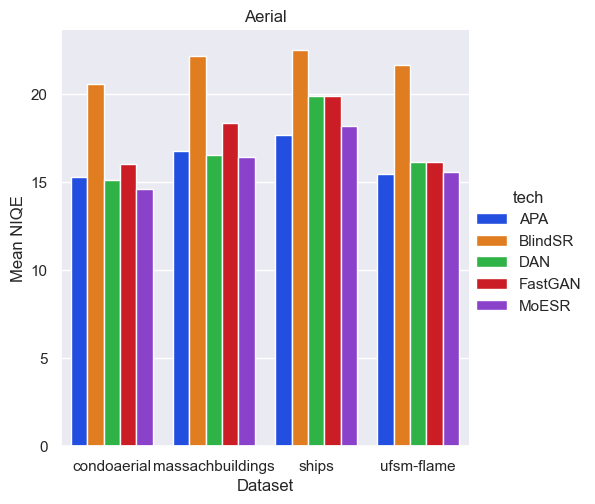} 
    \caption{Aerial} 
   \end{subfigure}
  \begin{subfigure}[b]{0.5\linewidth}
    \centering
    \includegraphics[width=1.0\linewidth]{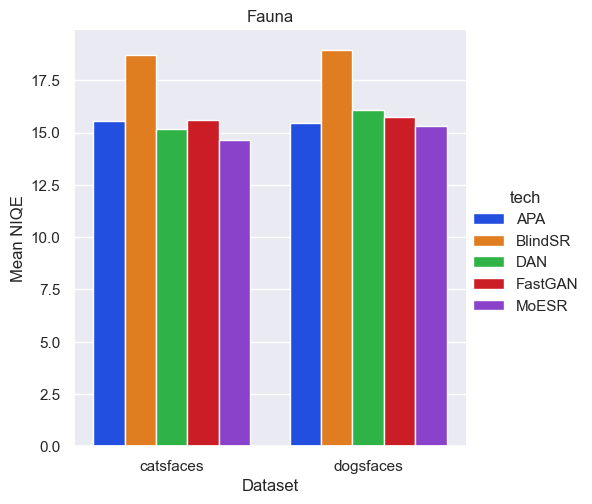} 
    \caption{Fauna} 
  \end{subfigure}

  \begin{subfigure}[b]{0.5\linewidth}
    \centering
    \includegraphics[width=1.0\linewidth]{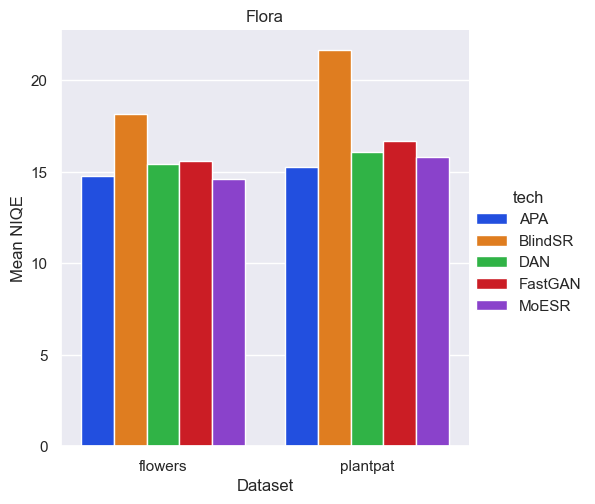} 
    \caption{Flora} 
   \end{subfigure} 
   \begin{subfigure}[b]{0.5\linewidth}
    \centering
    \includegraphics[width=1.0\linewidth]{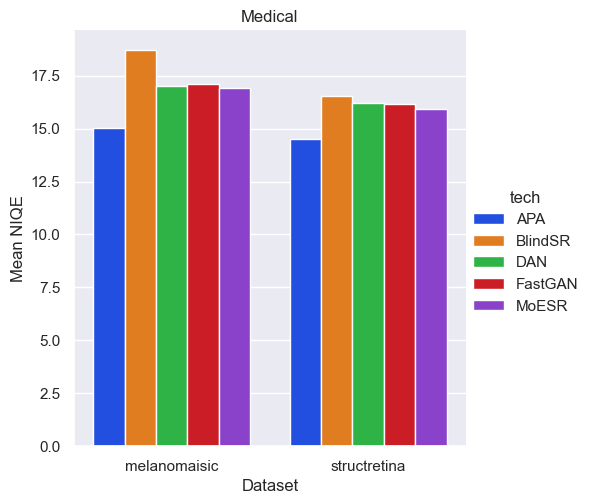} 
    \caption{Medical} 
  \end{subfigure} 
  
   \begin{subfigure}[b]{0.5\linewidth}
    \centering
    \includegraphics[width=1.0\linewidth]{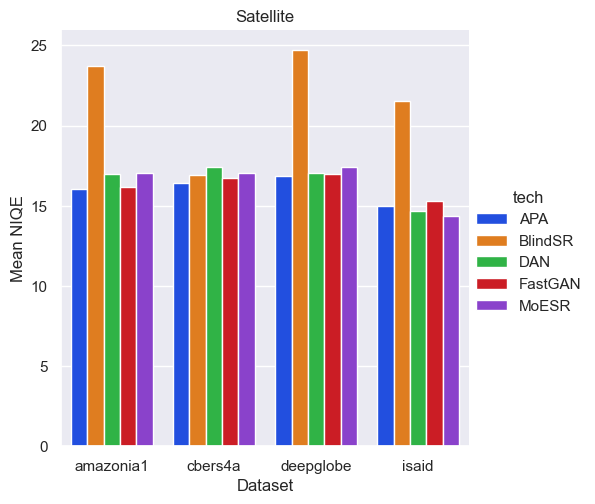} 
    \caption{Satellite} 
  \end{subfigure}

  \caption{Mean NIQE values: All broader domains and datasets}
  \label{niqefig} 
\end{figure}

Table \ref{niqedom} presents the NIQE values in the perspective of the broader domain. In this case, we consider the mean NIQE values for all the datasets of a certain domain where the best (minimum) and worst (maximum) values, and the techniques which produced them, are shown. On the other hand, Table \ref{niqeset} presents a finer analysis per dataset and also considering the mean NIQE values. Recall that as lower the NIQE value, the better the perceptual quality. Fig. \ref{niqefig} shows all mean NIQE values considering all broader domains and datasets.

\begin{table}[!htb]
\centering
\caption{Mean NIQE values: Broader domains}
\begin{tabular}{c|cc|cc}
\hline
\footnotesize \ngr{Domain} & \multicolumn{2}{c|}{\footnotesize \ngr{Min (Best)}}  & \multicolumn{2}{c}{\footnotesize \ngr{Max (Worst)}} \\ 
\hline
 
 & \footnotesize Technique & \footnotesize NIQE & \footnotesize Technique & \footnotesize NIQE   \\ 
 \hline
\footnotesize Aerial & \footnotesize MoESR & \footnotesize 14.648013 & \footnotesize  BlindSR & \footnotesize 22.534419  \\ 

\footnotesize Fauna & \footnotesize MoESR & \footnotesize 14.629656 & \footnotesize  BlindSR & \footnotesize 18.961343  \\ 

\footnotesize Flora & \footnotesize MoESR & \footnotesize 14.615975 & \footnotesize  BlindSR & \footnotesize 21.668490 \\

\footnotesize Medical & \footnotesize APA & \footnotesize 14.520624 & \footnotesize  BlindSR & \footnotesize 18.717661 \\

\footnotesize Satellite & \footnotesize MoESR & \footnotesize 14.385091 & \footnotesize  BlindSR & \footnotesize 24.726333 \\

\hline
\end{tabular}
\label{niqedom}
\end{table}

\begin{table}[!htb]
\centering
\caption{Mean NIQE values: Datasets}
\begin{tabular}{c|c|cc|cc}
\hline
\footnotesize \ngr{Domain} & \footnotesize \ngr{Dataset} & \multicolumn{2}{c|}{\footnotesize \ngr{Min (Best)}}  & \multicolumn{2}{c}{\footnotesize \ngr{Max (Worst)}} \\ 
\hline
 
\footnotesize Aerial & \footnotesize \maq{condoaerial} &  \footnotesize MoESR &  \footnotesize 14.648013  & \footnotesize BlindSR &  \footnotesize 20.611777 \\ 
 & \footnotesize   \maq{massachbuildings} & \footnotesize MoESR &  \footnotesize 16.427236  & \footnotesize BlindSR &  \footnotesize 22.199742 \\ 
 & \footnotesize   \maq{ships} & \footnotesize APA &  \footnotesize 17.688239  & \footnotesize BlindSR &  \footnotesize 22.534419 \\ 
 & \footnotesize   \maq{ufsm-flame} & \footnotesize APA &  \footnotesize 15.492423  & \footnotesize BlindSR &  \footnotesize 21.684649 \\   \hline
 \hline
 
 \footnotesize Fauna & \footnotesize \maq{catsfaces} & \footnotesize MoESR &  \footnotesize 14.629656  & \footnotesize BlindSR &  \footnotesize 18.702201 \\ 
 & \footnotesize   \maq{dogsfaces} & \footnotesize MoESR &  \footnotesize 15.314877  & \footnotesize BlindSR &  \footnotesize 18.961343 \\   \hline
 \hline
 
 \footnotesize Flora & \footnotesize \maq{flowers} & \footnotesize MoESR &  \footnotesize 14.615974  & \footnotesize BlindSR &  \footnotesize 18.174120 \\ 
 & \footnotesize  \maq{plantpat} & \footnotesize APA &  \footnotesize 15.241252     & \footnotesize BlindSR &  \footnotesize 21.668490 \\   \hline
 \hline
 
 \footnotesize Medical & \footnotesize \maq{melanomaisic} & \footnotesize APA &  \footnotesize 15.025945  & \footnotesize BlindSR &  \footnotesize 18.717661 \\ 
 & \footnotesize  \maq{structretina} & \footnotesize APA &  \footnotesize 14.520624  & \footnotesize BlindSR &  \footnotesize 16.517888 \\  \hline
 \hline
 
 \footnotesize Satellite & \footnotesize \maq{amazonia1} & \footnotesize APA &  \footnotesize 16.027069  & \footnotesize BlindSR &  \footnotesize 23.698020 \\ 
 & \footnotesize \maq{cbers4a} & \footnotesize APA &  \footnotesize 16.398990  & \footnotesize DAN &  \footnotesize 17.438634 \\ 
 & \footnotesize \maq{deepglobe} & \footnotesize APA &  \footnotesize 16.863639  & \footnotesize BlindSR &  \footnotesize 24.726333 \\ 
 & \footnotesize \maq{isaid} & \footnotesize MoESR &  \footnotesize 14.385091  & \footnotesize BlindSR &  \footnotesize 21.504252 \\ 
 \hline

\end{tabular}
\label{niqeset}
\end{table}


Results show that if we consider the broader domain (Table \ref{niqedom}), MoESR was the most outstanding approach being the best in four out of the five domains. Only in the Medical domain APA was the top technique. Moreover, BlindSR was the worst strategy with the maximum NIQE values for all domains. However, in the analysis per dataset (Table \ref{niqeset}), APA was the best technique obtaining the minimum NIQE values in eight out of 14 datasets. MoESR was the best in six out of them. BlindSR was, again, the worst of all the techniques.

Since MoESR was the best in one perspective and APA was the winner in the other analysis, we performed an improvement evaluation to reach a final decision in accordance with the NIQE value. In other words, the point is not only to say that a DL technique is better than the other but how much better it is. The improvement metric, $I\%$, is calculated as follows:

\begin{equation} \label{impeq}
I\% = \frac{(W - B) \times 100}{B}
\end{equation}

\noindent where $B$ and $W$ are the best and worst value of the metric, respectively, comparing both techniques under the same dataset. In Table \ref{niqeimmoe}, we show the improvement of MoESR over APA regarding NIQE where in all datasets in this table, MoESR was superior than APA. As highlighted in \ngr{bold}, the \maq{catsfaces} dataset was the one where MoESR got the highest improvement (6.472). In this case, $B = 14.629656$ and $W = 15.576498$. Hence, to know the APA's NIQE ($W$), we calculate:

\begin{equation} \label{impeq2}
W = B + \frac{I\%}{100} \times B.
\end{equation}  

The idea is that the improvement is the additional gain that the best approach has compared to the worst one. This is because as lower the NIQE value, the better.

\begin{table}[!htb]
\centering
\caption{Improvement of MoESR over APA}
\begin{tabular}{cccc}
\hline
\footnotesize \ngr{Dataset} & \footnotesize \ngr{MoESR}  & \footnotesize \ngr{APA}  & \footnotesize \ngr{$I\%$}\\ 
\hline
 
 \footnotesize \maq{condoaerial} & \footnotesize  14.648013 & \footnotesize  15.292783 & \footnotesize  4.402   \\ 
 
 \footnotesize \maq{massachbuildings} & \footnotesize  16.427236 & \footnotesize  16.760828 & \footnotesize 2.031 \\
 
 \footnotesize \maq{catsfaces} & \footnotesize  14.629656 & \footnotesize 15.576498 & \footnotesize \ngr{6.472} \\
  
 \footnotesize \maq{dogsfaces} & \footnotesize 15.314877 & \footnotesize 15.444758 & \footnotesize 0.848 \\
 
 \footnotesize \maq{flowers} & \footnotesize 14.615974 & \footnotesize 14.787114 & \footnotesize 1.171 \\
 
 \footnotesize \maq{isaid} & \footnotesize 14.385091 & \footnotesize 15.01201 & \footnotesize 4.358 \\ \\

\footnotesize $\overline{I\%}$ & \footnotesize  & \footnotesize  & \footnotesize 3.214 \\
  
\hline
\end{tabular}
\label{niqeimmoe}
\end{table}

In Table \ref{niqeimapa} is the other way around where we can see the improvement of APA over MoESR. The reasoning is the same as presented in Equations \ref{impeq} and \ref{impeq2}. Thus, the highest improvement (12.489) of APA over MoESR was obtained in the \maq{melanomaisic} dataset. Note that this improvement of APA is almost twice the best improvement of MoESR. Furthermore, the average improvement ($\overline{I\%}$) is also higher favouring APA over MoESR. Thus, based on the NIQE metric, the best approach was APA followed by MoESR. The worst was BlindSR.

\begin{table}[!htb]
\centering
\caption{Improvement of APA over MoESR}
\begin{tabular}{cccc}
\hline
\footnotesize \ngr{Dataset} & \footnotesize \ngr{APA}  & \footnotesize \ngr{MoESR}  & \footnotesize \ngr{$I\%$}\\ 
\hline
 
 \footnotesize \maq{ships} & \footnotesize   17.688239 & \footnotesize  18.220023 & \footnotesize  3.006   \\ 
 
 \footnotesize \maq{ufsm-flame} & \footnotesize  15.492423 & \footnotesize  15.614556 & \footnotesize 0.788 \\
 
 \footnotesize \maq{plantpat} & \footnotesize  15.241252 & \footnotesize 15.80596 & \footnotesize 3.705 \\
  
 \footnotesize \maq{melanomaisic} & \footnotesize  15.025945 & \footnotesize 16.902496 & \footnotesize  \ngr{12.489} \\
 
 \footnotesize \maq{structretina} & \footnotesize 14.520624 & \footnotesize 15.914241 & \footnotesize 9.598 \\
 
 \footnotesize \maq{amazonia1} & \footnotesize 16.027069 & \footnotesize 17.041395 & \footnotesize  6.329 \\
 
 \footnotesize \maq{cbers4a} & \footnotesize 16.39899 & \footnotesize 17.037916 & \footnotesize  3.896 \\
 
 \footnotesize \maq{deepglobe} & \footnotesize 16.863639 & \footnotesize 17.43862 & \footnotesize  3.410 \\ \\

\footnotesize $\overline{I\%}$ & \footnotesize  & \footnotesize  & \footnotesize 5.403 \\
  
\hline
\end{tabular}
\label{niqeimapa}
\end{table}

\subsubsection{MANIQA} \label{maniqaexp}

\begin{figure}[p] 
  \begin{subfigure}[b]{0.5\linewidth}
    \centering
    \includegraphics[width=1.0\linewidth]{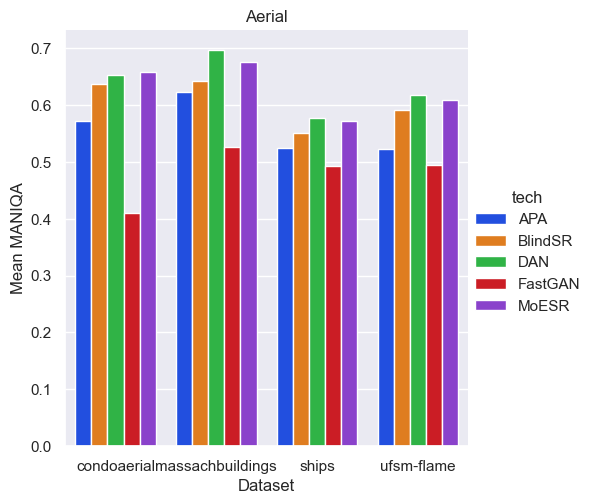} 
    \caption{Aerial} 
   \end{subfigure}
  \begin{subfigure}[b]{0.5\linewidth}
    \centering
    \includegraphics[width=1.0\linewidth]{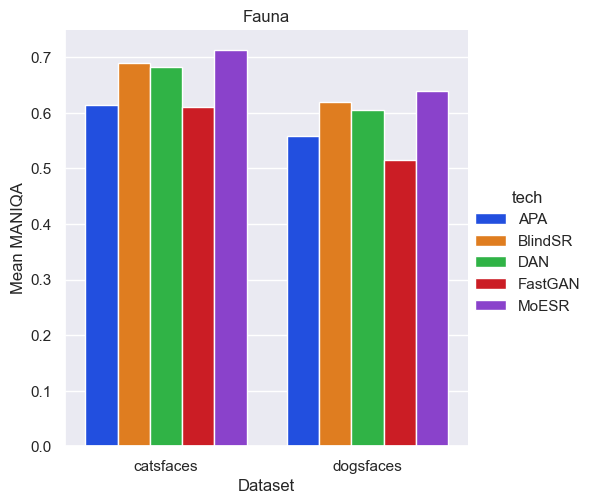} 
    \caption{Fauna} 
  \end{subfigure}

  \begin{subfigure}[b]{0.5\linewidth}
    \centering
    \includegraphics[width=1.0\linewidth]{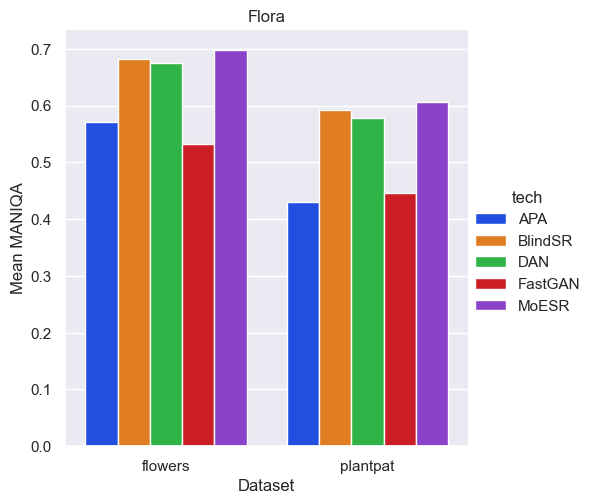} 
    \caption{Flora} 
   \end{subfigure} 
   \begin{subfigure}[b]{0.5\linewidth}
    \centering
    \includegraphics[width=1.0\linewidth]{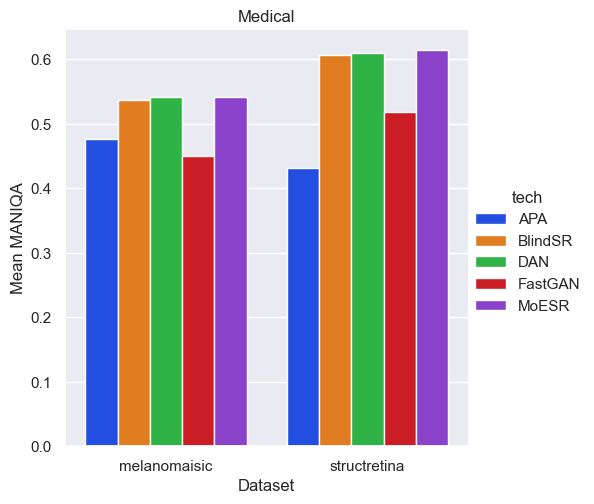} 
    \caption{Medical} 
  \end{subfigure} 
  
   \begin{subfigure}[b]{0.5\linewidth}
    \centering
    \includegraphics[width=1.0\linewidth]{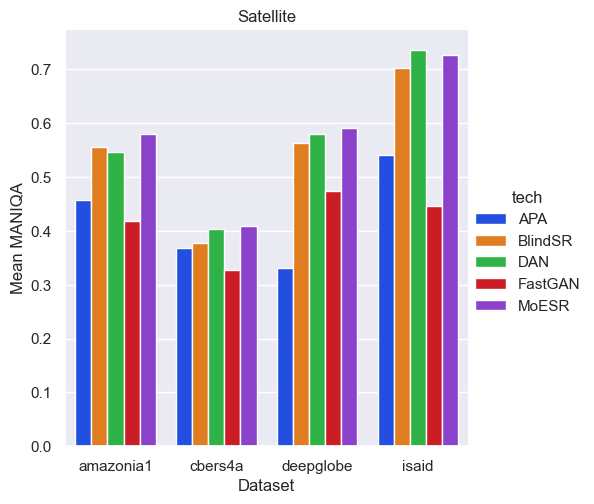} 
    \caption{Satellite} 
    \label{mansatellite}
  \end{subfigure}

  \caption{Mean MANIQA scores: All broader domains and datasets}
  \label{manfig} 
\end{figure}

In Table \ref{mandom}, the MANIQA scores are  shown in the perspective of the broader domain while Table \ref{manset} presents the per dataset analysis. As previously, we show the mean values of the metric. Recall that as higher the MANIQA score, the better the perceptual quality. Fig. \ref{manfig} shows all mean MANIQA scores considering all broader domains and datasets.

\begin{table}[!htb]
\centering
\caption{Mean MANIQA scores: Broader domains}
\begin{tabular}{c|cc|cc}
\hline
\footnotesize \ngr{Domain} & \multicolumn{2}{c|}{\footnotesize \ngr{Max (Best)}}  & \multicolumn{2}{c}{\footnotesize \ngr{Min (Worst)}} \\ 
\hline
 
 & \footnotesize Technique & \footnotesize MANIQA & \footnotesize Technique & \footnotesize MANIQA   \\ 
 \hline
\footnotesize Aerial & \footnotesize DAN & \footnotesize 0.696858 & \footnotesize  FastGAN & \footnotesize 0.409388  \\ 

\footnotesize Fauna & \footnotesize MoESR & \footnotesize 0.713253 & \footnotesize  FastGAN & \footnotesize 0.515693   \\ 

\footnotesize Flora & \footnotesize MoESR & \footnotesize 0.698373 & \footnotesize  APA & \footnotesize 0.430053  \\

\footnotesize Medical & \footnotesize MoESR & \footnotesize 0.614705 & \footnotesize  APA & \footnotesize 0.432007  \\

\footnotesize Satellite & \footnotesize DAN & \footnotesize 0.736443 & \footnotesize  FastGAN & \footnotesize 0.327089  \\

\hline
\end{tabular}
\label{mandom}
\end{table}

\begin{table}[!htb]
\centering
\caption{Mean MANIQA scores: Datasets}
\begin{tabular}{c|c|cc|cc}
\hline
\footnotesize \ngr{Domain} & \footnotesize \ngr{Dataset} & \multicolumn{2}{c|}{\footnotesize \ngr{Max (Best)}}  & \multicolumn{2}{c}{\footnotesize \ngr{Min (Worst)}} \\ 
\hline
 
\footnotesize Aerial & \footnotesize \maq{condoaerial} &  \footnotesize MoESR &  \footnotesize 0.657257  & \footnotesize FastGAN &  \footnotesize 0.409388 \\ 
 & \footnotesize   \maq{massachbuildings} & \footnotesize DAN &  \footnotesize 0.696858  & \footnotesize FastGAN &  \footnotesize 0.525381 \\ 
 & \footnotesize   \maq{ships} & \footnotesize DAN &  \footnotesize 0.577708  & \footnotesize FastGAN &  \footnotesize 0.492311  \\ 
 & \footnotesize   \maq{ufsm-flame} & \footnotesize DAN &  \footnotesize 0.618042  & \footnotesize FastGAN &  \footnotesize 0.493545   \\   \hline
 \hline
 
 \footnotesize Fauna & \footnotesize \maq{catsfaces} & \footnotesize MoESR &  \footnotesize 0.713253  & \footnotesize FastGAN &  \footnotesize 0.611105 \\ 
 & \footnotesize   \maq{dogsfaces} & \footnotesize MoESR &  \footnotesize 0.638982  & \footnotesize FastGAN &  \footnotesize 0.515693 \\   \hline
 \hline
 
 \footnotesize Flora & \footnotesize \maq{flowers} & \footnotesize MoESR &  \footnotesize 0.698373  & \footnotesize FastGAN &  \footnotesize 0.533008 \\ 
 & \footnotesize  \maq{plantpat} & \footnotesize MoESR &  \footnotesize 0.606683     & \footnotesize APA &  \footnotesize 0.430053  \\   \hline
 \hline
 
 \footnotesize Medical & \footnotesize \maq{melanomaisic} & \footnotesize DAN &  \footnotesize 0.542052  & \footnotesize FastGAN &  \footnotesize 0.449874     \\ 
 & \footnotesize  \maq{structretina} & \footnotesize MoESR &  \footnotesize  0.614705 & \footnotesize APA &  \footnotesize 0.432007 \\  \hline
 \hline
 
 \footnotesize Satellite & \footnotesize \maq{amazonia1} & \footnotesize MoESR &  \footnotesize 0.579970  & \footnotesize FastGAN &  \footnotesize 0.417986 \\ 
 & \footnotesize \maq{cbers4a} & \footnotesize MoESR &  \footnotesize 0.408773  & \footnotesize FastGAN &  \footnotesize 0.327089  \\ 
 & \footnotesize \maq{deepglobe} & \footnotesize MoESR &  \footnotesize 0.591723  & \footnotesize APA &  \footnotesize 0.330667   \\ 
 & \footnotesize \maq{isaid} & \footnotesize DAN &  \footnotesize 0.736443  & \footnotesize FastGAN &  \footnotesize 0.445917 \\ 
 \hline

\end{tabular}
\label{manset}
\end{table}

In the broader domain perspective, MoESR was the best (three out five top results) followed by DAN (two best positions). Unlike the NIQE metric where APA was considered the best DL technique overall, it performed poorly in accordance with the MANIQA score occupying the penultimate place (in two domains, Flora and Medical, APA was the worst approach). FastGAN was the worst of all the techniques. Again unlike the results presented in the previous section, here we have a completely agreement between the broader domain and the per dataset perspectives. Thus, considering each dataset in its own, MoESR got again the top place (nine best MANIQA scores) while DAN was the second best technique (five best MANIQA scores). Likewise, APA got the penultimate place (three worst MANIQA scores) and FastGAN was the worst of all the DL techniques.

Despite such an agreement, we accomplished the improvement analysis to check these results considering the two best DL models: MoESR and DAN. But, we slightly changed the way to calculate the improvement metric as shown below:

\begin{equation} \label{impeq3}
I\% = \frac{(B - W) \times 100}{W}
\end{equation}

\noindent where $B$ and $W$ are the best and worst value of the metric, respectively, comparing both techniques under the same dataset. And now, the idea is that the improvement is the additional gain that the worst approach needs to reach the best one. This is because as higher the MANIQA score, the better. Thus, we use the following equation:

\begin{equation} \label{impeq4}
B = W + \frac{I\%}{100} \times W.
\end{equation}

Table \ref{manimdan} presents the improvement of DAN over MoESR while Table \ref{manimmoe} shows the opposite (MoESR over DAN). As we can see in \ngr{bold} in both tables, the highest improvement of MoESR (6.069) is almost twice the highest due to DAN (3.245). Moreover, MoESR's average improvement is also better than the DAN's average improvement. These results confirms the previous conclusions and MoESR was the best technique followed by DAN.

\begin{table}[!htb]
\centering
\caption{Improvement of DAN over MoESR}
\begin{tabular}{cccc}
\hline
\footnotesize \ngr{Dataset} & \footnotesize \ngr{DAN}  & \footnotesize \ngr{MoESR}  & \footnotesize \ngr{$I\%$}\\ 
\hline
 
 \footnotesize \maq{massachbuildings} & \footnotesize  0.696858 & \footnotesize  0.674955 & \footnotesize  \ngr{3.245}   \\ 
 
 \footnotesize \maq{ships} & \footnotesize  0.577708 & \footnotesize  0.571233 & \footnotesize 1.134 \\
 
 \footnotesize \maq{ufsm-flame} & \footnotesize  0.618042 & \footnotesize 0.608577 & \footnotesize 1.555 \\
  
 \footnotesize \maq{melanomaisic} & \footnotesize 0.542052 & \footnotesize 0.541265 & \footnotesize 0.145 \\
 
\footnotesize \maq{isaid} & \footnotesize 0.736443 & \footnotesize 0.72679 & \footnotesize 1.328 \\ \\

\footnotesize $\overline{I\%}$ & \footnotesize  & \footnotesize  & \footnotesize 1.481 \\
  
\hline
\end{tabular}
\label{manimdan}
\end{table}

\begin{table}[!htb]
\centering
\caption{Improvement of MoESR over DAN}
\begin{tabular}{cccc}
\hline
\footnotesize \ngr{Dataset} & \footnotesize \ngr{MoESR}  & \footnotesize \ngr{DAN}  & \footnotesize \ngr{$I\%$}\\ 
\hline
 
 \footnotesize \maq{condoaerial} & \footnotesize  0.657257 & \footnotesize 0.653125 & \footnotesize  0.633   \\ 
 
 \footnotesize \maq{catsfaces} & \footnotesize  0.713253 & \footnotesize 0.681594 & \footnotesize 4.645 \\
  
 \footnotesize \maq{dogsfaces} & \footnotesize 0.638982 & \footnotesize 0.605406 & \footnotesize 5.546 \\
 
 \footnotesize \maq{flowers} & \footnotesize 0.698373 & \footnotesize 0.674927 & \footnotesize 3.474 \\
 
 \footnotesize \maq{plantpat} & \footnotesize 0.606683 & \footnotesize 0.578225 & \footnotesize 4.922 \\ 
 
 \footnotesize \maq{structretina} & \footnotesize 0.614705 & \footnotesize 0.609501 & \footnotesize 0.854 \\ 
 
  \footnotesize \maq{amazonia1} & \footnotesize 0.57997 & \footnotesize 0.546786 & \footnotesize \ngr{6.069} \\ 
  
  \footnotesize \maq{cbers4a} & \footnotesize 0.408773 & \footnotesize 0.403834 & \footnotesize 1.223 \\ 
  
   \footnotesize \maq{deepglobe} & \footnotesize 0.591723 & \footnotesize 0.580666 & \footnotesize 1.904 \\  \\

\footnotesize $\overline{I\%}$ & \footnotesize  & \footnotesize  & \footnotesize 3.252 \\
  
\hline
\end{tabular}
\label{manimmoe}
\end{table}

Considering both metrics, NIQE and MANIQA score, we can state that MoESR was the most outstanding approach. It presented a consistent performance under NIQE, being the second best technique, and, as we have just said, it got the top place based on the MANIQA score. Note that we saw contradictory performances regarding APA where it was the best strategy evaluated via NIQE and almost the worst approach, if we take into account the MANIQA score.


\subsection{RQ\_2}

Question \ngr{RQ\_2} is about the similarity of behaviours between the two best approaches according to the MANIQA score: MoESR and DAN. Here, our intention is to see whether the two best models “agree” in terms of the HR images they create. Thus, we took the 10 images with the best MANIQA scores from both techniques, for each dataset, and generated two sets $High_M$ and $High_D$, meaning the scores of the images from MoESR and DAN, respectively.  Hence, we just calculated the cardinality of a new set ($C_B$) derived by the intersection between the previous two sets:

\begin{equation} \label{eqH}
 \lvert C_B \rvert =  \lvert High_M \cap High_D \rvert.
\end{equation}
 
 We also did the same and calculated the cardinality of the set of non-common elements, $N_B$, such that:
 \begin{equation} \label{eqL}
  \lvert N_B \rvert =  \frac{\lvert High_M \cup High_D \rvert - \lvert High_M \cap High_D \rvert}{2}. 
  \end{equation}
  
For instance, for the \maq{condoaerial} dataset,  $\lvert C_B(condoaerial) \rvert = 7$. This implies that seven out of the 10 images with the highest MANIQA scores are common to MoESR and DAN, and hence there is an agreement between both techniques regarding these images. Thus, $\lvert N_B(condoaerial) \rvert = 3$, representing that three out of the 10 images are not common showing disagreement. The average value of the cardinalities of all common sets is $\overline{C_B(all)} = 7.36$ while for the non-common case is $\overline{N_B(all)} = 2.64$.

We did the same with the 10 images with lowest scores and created the sets $Low_M$ and $Low_D$ due to MoESR and DAN, respectively. Likewise, we calculated the cardinality of $C_W$ and $N_W$ for each dataset, just replacing the high sets by the low ones in the Equations \ref{eqH} and \ref{eqL}. For example, for the \maq{amazonia1} dataset, $\lvert C_W(amazonia1) \rvert = 6$ and $\lvert N_W(amazonia1) \rvert = 4$, meaning that six images with the lowest scores are common and four are not common. The average value of the cardinalities of all common sets is $\overline{C_W(all)} = 7.5$ while for the non-common case is $\overline{N_W(all)} = 2.5$.  

\begin{figure}[!htb] 
  \begin{subfigure}[b]{0.5\linewidth}
    \centering
    \includegraphics[width=1.0\linewidth]{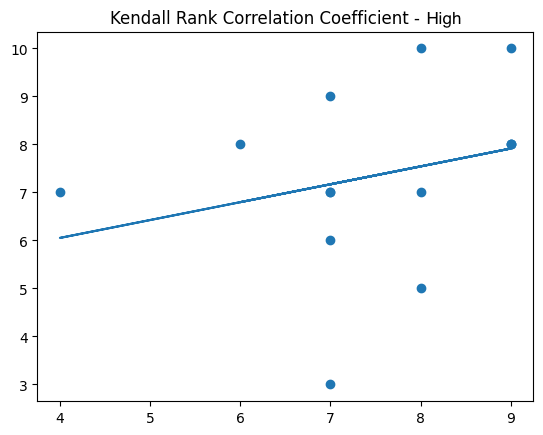} 
    \caption{Correlation of $C$ sets} 
    \label{kenH} 
   \end{subfigure}
  \begin{subfigure}[b]{0.5\linewidth}
    \centering
    \includegraphics[width=1.0\linewidth]{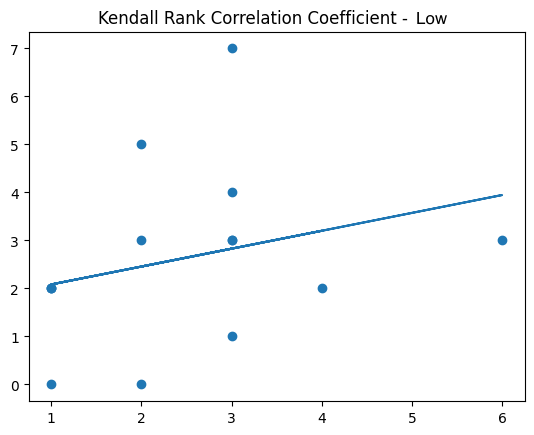} 
    \caption{Correlation of $N$ sets} 
    \label{kenL} 
 \end{subfigure} 
      
  \caption{Kendall Rank Correlation Coefficient: results}
  \label{kendall} 
\end{figure}

Finally, we obtained the Kendall Rank Correlation Coefficient (Kendall's $\tau$ coefficient) for two situations. The first case is considering $C_B$ and $C_W$ and the second situation is based on $N_B$ and $N_W$. In both cases, $\tau = 0.306912$ showing a good correlation between each pair of variables. Fig. \ref{kendall} shows the correlation in detail.

The interpretation of these results is that the images detected as having the best, as well as the worst, perceptual qualities, based on the MANIQA scores, are somewhat ``common'' to both techniques. Hence, we can conclude that both approaches present a similar behaviour. It is important to emphasise that the relevance of this analysis is to realise whether the two best approaches generate HR images with better or worse perception qualities in a relatively uniform manner, taking the LR images as input.

\section{Discussion} \label{disc}

\begin{figure}[!htb]
\centering
\includegraphics[width=1.0\textwidth]{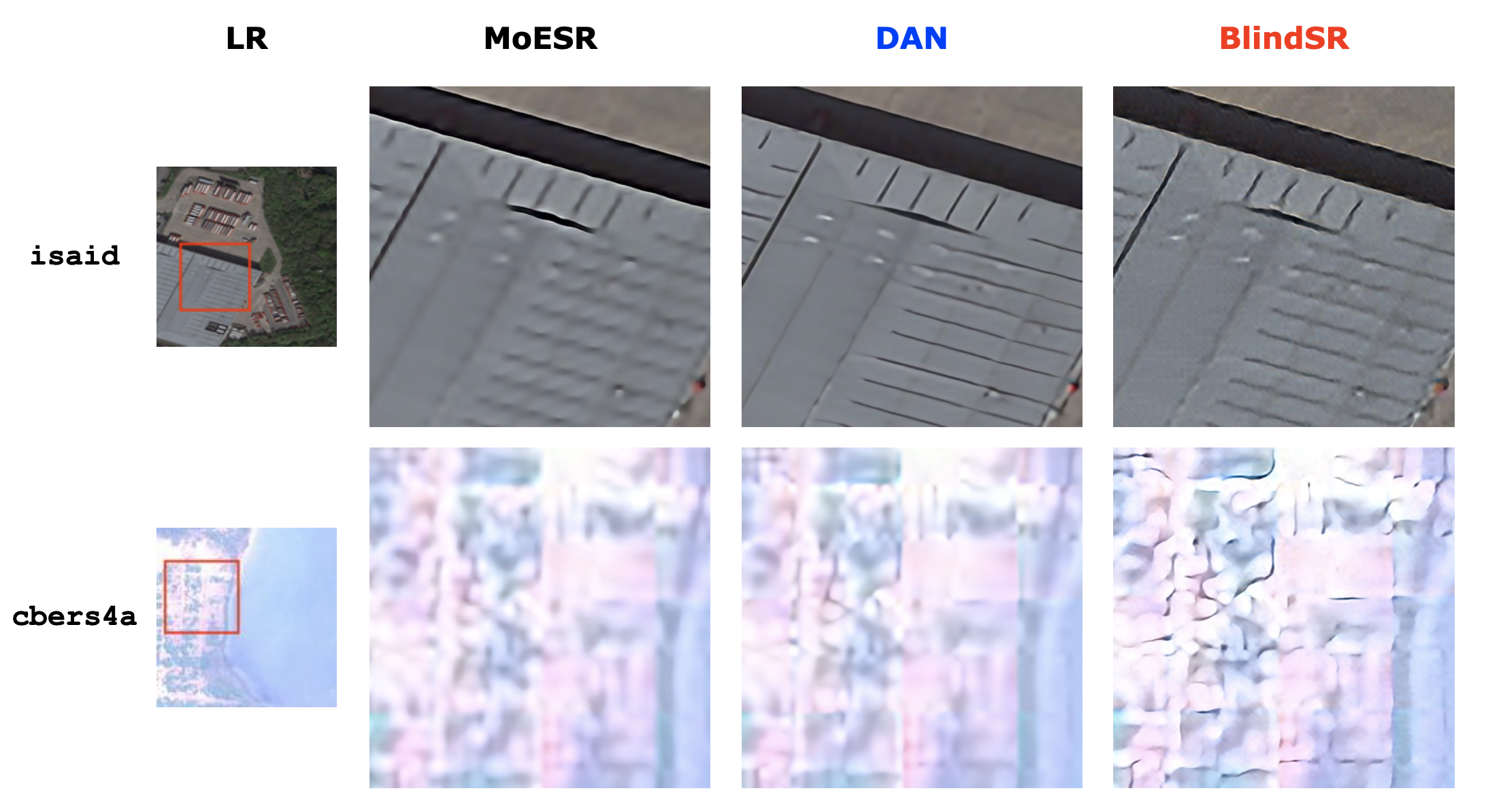}
\caption{Visual analysis: highest and lowest MANIQA scores}
\label{vis1}
\end{figure}

We start this section by stressing some points based on a visual analysis. Fig. \ref{vis1} shows the LR image and a piece of the corresponding HR image generated by the three single-image blind SR techniques. The top HR image generated by DAN, based on an LR image of the \maq{isaid} Satellite dataset, is the one with the highest MANIQA score (0.850084) of all images. We clearly see that the roof divisions of the building appear more vivid in the image created by DAN. MoESR's image appears blurred while BlindSR's image seems to present something like a salt-and-pepper noise. 

On the other hand, the bottom HR image created by BlindSR is the one with the lowest MANIQA score (0.240234) of all images. The LR image is from the \maq{cbers4a} Satellite dataset where the scene presents water and land. This is a cloudy scene obtained by the CBERS-4A satellite and it is interesting to realise that such bright images from \maq{cbers4a} in general produced the lowest MANIQA scores (see Fig. \ref{mansatellite}). 

\begin{figure}[!htb]
\centering
 
 {\begin{tabular}{lcccc} 
      \footnotesize \ngr{Dataset} &  \footnotesize \ngr{LR} &  \footnotesize \ngr{MoESR} &  \footnotesize \ngr{DAN} &  \footnotesize \ngr{BlindSR} \\
      
      \footnotesize \maq{condoaerial} &  \parbox[c]{6em}{ \includegraphics[width=2cm]{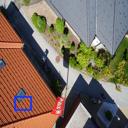}} & 
      \parbox[c]{6em}{ \includegraphics[width=2cm]{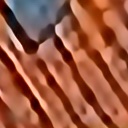}} & 
      \parbox[c]{6em}{ \includegraphics[width=2cm]{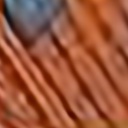}} &
      \parbox[c]{6em}{ \includegraphics[width=2cm]{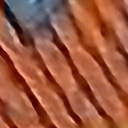}} \vspace{0.1cm} \\
      
      \footnotesize \maq{flowers} &  \parbox[c]{6em}{\includegraphics[width=2cm]{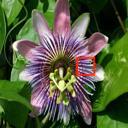}} & 
      \parbox[c]{6em}{\includegraphics[width=2cm]{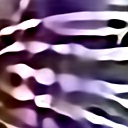}} &
      \parbox[c]{6em}{\includegraphics[width=2cm]{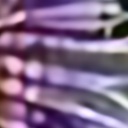}} &
      \parbox[c]{6em}{ \includegraphics[width=2cm]{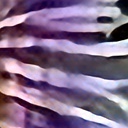}} \vspace{0.1cm} \\
      
      \footnotesize \maq{melanomaisic} &  \parbox[c]{6em}{\includegraphics[width=2cm]{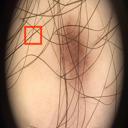}} & 
      \parbox[c]{6em}{\includegraphics[width=2cm]{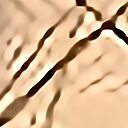}} &
      \parbox[c]{6em}{\includegraphics[width=2cm]{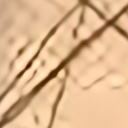}} &
      \parbox[c]{6em}{ \includegraphics[width=2cm]{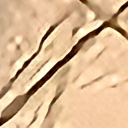}} \vspace{0.1cm} \\
      
       \footnotesize \maq{amazonia1} &  \parbox[c]{6em}{\includegraphics[width=2cm]{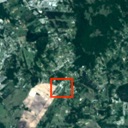}} & 
      \parbox[c]{6em}{\includegraphics[width=2cm]{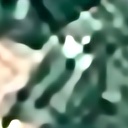}} &
      \parbox[c]{6em}{\includegraphics[width=2cm]{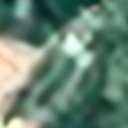}} &
      \parbox[c]{6em}{ \includegraphics[width=2cm]{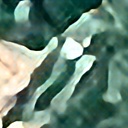}} \vspace{0.1cm} \\
      
       \footnotesize \maq{deepglobe} &  \parbox[c]{6em}{\includegraphics[width=2cm]{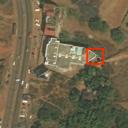}} & 
      \parbox[c]{6em}{\includegraphics[width=2cm]{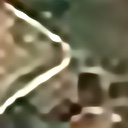}} &
      \parbox[c]{6em}{\includegraphics[width=2cm]{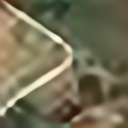}} &
      \parbox[c]{6em}{ \includegraphics[width=2cm]{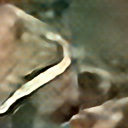}} \vspace{0.1cm} \\

      \end{tabular}}

\caption{Other created HR images by the single-image blind SR approaches} 
 \label{vis2}
 \end{figure}

Other LR images and the corresponding piece of HR images created by the three single-image blind SR models are presented in Fig. \ref{vis2}. MoESR produced the highest MANIQA score for the top images (\maq{condoaerial}, \maq{flowers}, \maq{melanomaisic}, \maq{amazonia1}) while DAN produced the highest score for the bottom one ({\maq{deepglobe}}). 

Since APA and FastGAN are GAN-based non-single image techniques, some known issues of GANs appeared here. In order to use APA in a custom dataset, three phases are required: prepare the dataset, training, and inference for generating images. As usual, the training is the more demanded step and APA indeed exceeded the limit we set up (four days) not finishing the training. We considered its latest model for inference. We used 4 x NVIDIA Volta V100 GPUs and the datasets are very small (100 samples). It is important to mention that other DL techniques demanded only one out of the four GPUs and the limit was enough for them. Thus, APA is a very ``heavy'' model. Furthermore, some HR images were flipped and others upside down compared to the source LR images. There are even cases where the images were not indeed generated (it is likely that the images are basically noise). 

\begin{figure}[!htb]
\centering
 
 {\begin{tabular}{cccc} 

      \footnotesize LR &  \footnotesize APA - Flipping &  \footnotesize  LR &  \footnotesize APA - Flipping \\

      \parbox[c]{6em}{ \includegraphics[width=2cm]{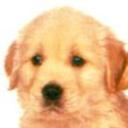}} &  
      \parbox[c]{6em}{ \includegraphics[width=2cm]{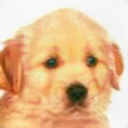}} & 
      \parbox[c]{6em}{ \includegraphics[width=2cm]{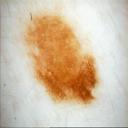}} &
      \parbox[c]{6em}{ \includegraphics[width=2cm]{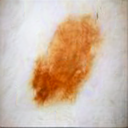}} \vspace{0.1cm} \\
      
      \footnotesize LR &  \footnotesize APA - Up Down &  \footnotesize  LR &  \footnotesize APA - Noise \\
      
      \parbox[c]{6em}{ \includegraphics[width=2cm]{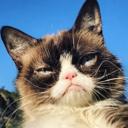}} &  
      \parbox[c]{6em}{ \includegraphics[width=2cm]{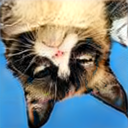}} & 
      \parbox[c]{6em}{ \includegraphics[width=2cm]{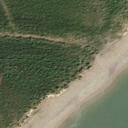}} &
      \parbox[c]{6em}{ \includegraphics[width=2cm]{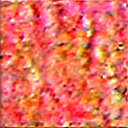}} \vspace{0.1cm} \\

      \end{tabular}}

\caption{Some issues presented by APA. Top row, left to right: \maq{dogsfaces}, \maq{melanomaisic}; Bottom row, left to right: \maq{catsfaces}, \maq{deepglobe}} 
 \label{apaissues}
 \end{figure}

Fig. \ref{apaissues} shows these cases where we have the LR image and the corresponding APA's HR image. Note that we downsampled the APA images to 128 x 128 pixels for this figure to compare to the LR images. Moreover, the \maq{deepglobe} image is here just to show a sample of the dataset since APA's output seems to be basically noise.

\begin{figure}[!htb]
\centering
{ \begin{tabular}{cccc}

       \parbox[c]{6em}{ \includegraphics[width=2cm]{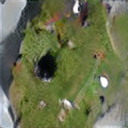}} &  
      \parbox[c]{6em}{ \includegraphics[width=2cm]{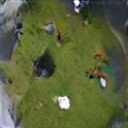}} & 
      \parbox[c]{6em}{ \includegraphics[width=2cm]{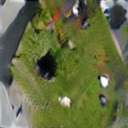}} &
      \parbox[c]{6em}{ \includegraphics[width=2cm]{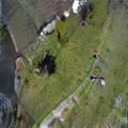}} \vspace{0.1cm} \\
      
       \parbox[c]{6em}{ \includegraphics[width=2cm]{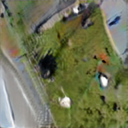}} &  
      \parbox[c]{6em}{ \includegraphics[width=2cm]{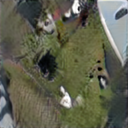}} & 
      \parbox[c]{6em}{ \includegraphics[width=2cm]{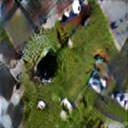}} &
      \parbox[c]{6em}{ \includegraphics[width=2cm]{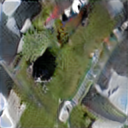}} \vspace{0.1cm} \\

      \end{tabular}}

\caption{Mode collapse in FastGAN: \maq{condoaerial}} 
 \label{fganissues}
 \end{figure}

As for FastGAN, the required phases to use it are training and inference, and its training is considerably faster/lighter than APA. But both FastGAN and APA presented a very known problem related to GANs: mode collapse \cite{mcgan/17}. It happens when the generator model produces a not very significant set of images that fail to capture the full diversity of the real data distribution. Thus, the fake created samples are quite similar or even identical. In Fig. \ref{fganissues}, we see some images from FastGAN where mode collapse occurs. Again, we downgraded the HR images for better presentation.

Based on the MANIQA scores (Section \ref{maniqaexp}), APA and FastGAN were the worst techniques. Thus, we can conclude based on the results of our experiment that, for blind image SR, single-image and non-GAN-based approaches are the best way to go. 

\begin{figure}[!htb] 
     \centering
    \includegraphics[width=1.0\linewidth]{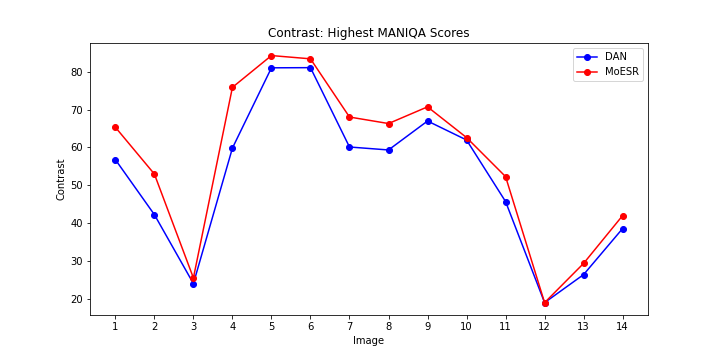} 
    \caption{Highest MANIQA scores: Contrast} 
    \label{highcon} 
\end{figure}

\begin{figure}[!htb] 
     \centering
    \includegraphics[width=1.0\linewidth]{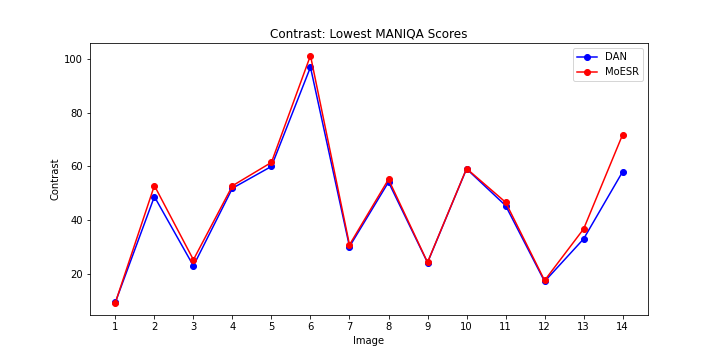} 
    \caption{Lowest MANIQA scores: Contrast} 
    \label{lowcon} 
\end{figure}

Considering the best two DL techniques in accordance with the MANIQA score, we may say that the HR images created by MoESR are sharper than the ones of DAN. Fig. \ref{highcon} presents the measures of the maximum overall contrast of the images with the highest MANIQA scores taking into account both techniques and the 14 datasets. For instance, for the \maq{amazonia1} dataset, the image derived by MoESR has higher MANIQA score than the one of DAN and, thus, we considered the former and the corresponding DAN's image. As for \maq{isaid}, DAN's image got higher MANIQA score and thus this image and the corresponding MoESR's image were analysed. Fig. \ref{lowcon} follows the same reasoning but now the images are the ones with the lowest MANIQA scores.  

Since the higher the maximum overall contrast the sharper the image, we can realise that the images of MoESR are sharper than the ones of DAN. But, note that perceptual quality is not necessarily the same thing as sharpness. An image with higher contrast does not imply that it has a better perceptual quality than one with lower contrast. 

A possible explanation for the images created by MoESR being sharper than DAN's images, and sometimes present oversharpening, is the combination of the perfomances of ISE and KEN. The ISE component is trained to detect blurry or oversharpened regions and predicts errors from the ground-truth image. Since KEN uses the sharpness measures from ISE to estimate the kernel and select the best pretrained model, misleading evaluations of sharpness by ISE may compromise the decision made by the KEN component.

On the other hand, HR images generated by DAN are generally blurry which might lead to a poor perceptual quality. In DAN, the kernel is initialised by Dirac function, and it is also reshaped and then reduced by principal component analysis (PCA) \cite{pcaref/93}. The kernel is reduced by PCA and thus the Estimator only needs to estimate the PCA result of the blur kernel. There is naturally loss of information when using PCA for dimensionality reduction, and recent evaluations show that PCA results are not as reliable and robust as it is usually assumed to be \cite{pcais/22}. This is a possible explanation for this issue related to DAN.

As mentioned in Section \ref{results}, there is not an agreement between the ranking of best and worst techniques considering a classical metric (NIQE) and a DNN-based one (MANIQA score). While APA presented the better performance under NIQE, it is almost the worst solution according to the MANIQA score. Thus, we believe that relying on a more recent NR-IQA metric, like the MANIQA score and which presented quite superior performance \cite{ntire/22} than traditional metrics, is more advisable. 

Notice that we do not have the mean opinion scores (MOSs) of the HR images as we do not submit them for observers to assign it. But, looking at the HR images generated by all DL techniques for all sets (see the datasets repository of this research), it is not difficult to see that the perception quality of the images as a whole needs to improve. Note that in MoESR and DAN, the generation of the HR images had to be done in two stages, using the 2x and 4x scaling factors in sequence. In general, there are several blurred images and others with oversharpening. Despite the sophistication of the evaluated DNNs, we believe that new approaches, addressing larger scaling factors, are necessary for the future.

\section{Conclusions} \label{conc}

Given the significant number of DL techniques/DNNs which are created on a daily basis, it is relevant to perform independent and unbiased controlled experiments to suggest the most suitable approaches for professionals. This article is in line with this reasoning, where we presented an large scaling factor (8x) experiment which evaluated recent DL techniques tailored for blind image SR: APA, BlindSR, DAN, FastGAN, and MoESR. Some of the techniques were designed for single-image while others for non-single-image blind SR. In addition to selecting a larger scaling factor, another difference in this study is that we showed a more detailed analysis of the results, also focusing on explaining in depth the behaviours of the approaches. Mostly based on publicly released sources of images, we adapted and created 14 small (100 samples) LR images datasets from five different broader domains to provide a significant range of distinct images to the techniques. We also considered two NR-IQA metrics, the classical NIQE and the DNN-based MANIQA score. 

Results show that the MoESR model was the most outstanding approach followed by DAN. GAN-based approaches presented classical issues such as mode collapse and noise, in some cases. The outcomes of our experiment suggest that single-image and non-GAN-based techniques are more promising for blind image SR, although such a conclusion needs more experimentation to be completely confirmed. By visually inspecting the HR images created by the techniques, we may state that, despite the remarkable solutions, new strategies are required for larger scaling factor requirements.

As future directions, we firstly intend to carry out even higher scaling factor experiments, such as 16x. We will also increase not only the amount of blind image SR approaches to assess but also the number of datasets to get more evidences about the conclusions we have made. Other recent NR-IQA metrics will also be considered in these new controlled experiments.




\section*{Acknowledgements}
This research was developed within  the project \ita{Classifica{\c c}{\~a}o de imagens via redes neurais profundas e grandes bases de dados para aplica{\c c}{\~o}es aeroespaciais} (Image classification via Deep neural networks and large databases for aeroSpace applications - \textbf{IDeepS}). The \textbf{IDeepS} ({\href{https://github.com/vsantjr/IDeepS}{https://github.com/vsantjr/IDeepS}) project is supported by the \ita{Laborat{\'o}rio Nacional de Computa{\c c}{\~a}o Cient{\'i}fica} (LNCC/MCTI, Brazil) via resources of the SDumont supercomputer (\href{http://sdumont.lncc.br}{http://sdumont.lncc.br}).

\balance
\bibliographystyle{unsrt}
\bibliography{BlindSRSantiagoJr}

\end{document}